\documentclass[%
 reprint,
superscriptaddress,
 amsmath,amssymb,
 aps,
prb,
]{revtex4-1}

\usepackage{graphicx}
\usepackage{dcolumn}
\usepackage{bm}
\usepackage{hyperref}

\usepackage{xcolor}

\begin{document}

\title{Control of field- and current-driven magnetic domain wall motion by exchange-bias in \texorpdfstring{C\lowercase{r}$_2$O$_3$/C\lowercase{o}/P\lowercase{t}}{} trilayers}

\author{B. J.~Jacot}
\affiliation{Department of Materials, ETH Zurich, CH-8093 Zurich, Switzerland}
\author{S.~Vélez}
\affiliation{Department of Materials, ETH Zurich, CH-8093 Zurich, Switzerland}
\affiliation{Condensed Matter Physics Center (IFIMAC), Instituto Nicolás Cabrera, and Departamento de Física de la Materia Condensada, Universidad Autónoma de Madrid, 28049 Madrid, Spain}
\author{P.~Noël}
\affiliation{Department of Materials, ETH Zurich, CH-8093 Zurich, Switzerland}
\author{P.~Helbingk}
\affiliation{Department of Materials, ETH Zurich, CH-8093 Zurich, Switzerland}
\author{F.~Binda}
\affiliation{Department of Materials, ETH Zurich, CH-8093 Zurich, Switzerland}
\author{C.-H.~Lambert}
\affiliation{Department of Materials, ETH Zurich, CH-8093 Zurich, Switzerland}
\author{P.~Gambardella}
\affiliation{Department of Materials, ETH Zurich, CH-8093 Zurich, Switzerland}

\date{September 27, 2022}

\begin{abstract}
We investigate the motion of magnetic domain walls driven by magnetic fields and current-driven spin-orbit torques in an exchange-biased system with perpendicular magnetization. We consider Cr$_2$O$_3$/Co/Pt trilayers as model system, in which the magnetization of the Co layer can be exchanged-biased out-of-plane or in-plane depending on the field cooling direction. In field-driven experiments, the in-plane exchange bias favors the propagation of the domain walls with internal magnetization parallel to the exchange bias field. In current-driven experiments, the domain walls propagate along the current direction, but the domain wall velocity increases and decreases symmetrically (anti-symmetrically) for both current polarities when the exchange bias is parallel (perpendicular) to the current line. At zero external field, the exchange bias modifies the velocity of current-driven domain wall motion by a factor of ten. We also find that the exchange bias remains stable under external fields up to 15~kOe and ns-long current pulses with current density up to 3.5$\times10^{12}$~A/m. Our results demonstrate versatile control of the domain wall motion by exchange bias, which is relevant to achieve field-free switching of the magnetization in perpendicular systems and current-driven manipulation of domain walls velocity in spintronic devices.
\end{abstract}

\maketitle

\section{\label{sec-intro} Introduction}

\begin{figure*}
\centering
\includegraphics[width=0.99\linewidth]{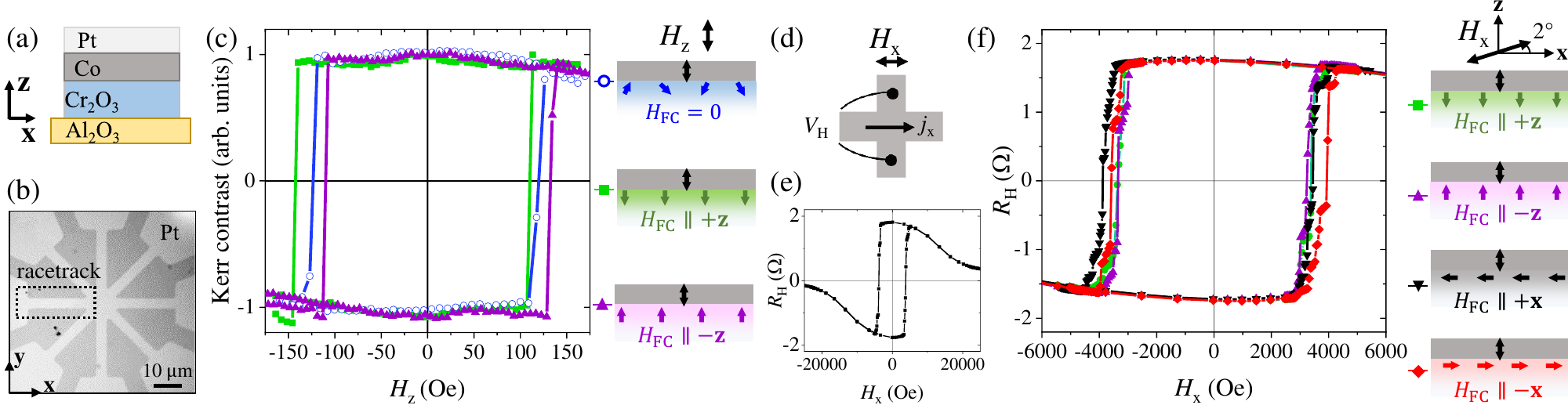}
\caption{\label{fig:hysteresis-loops}
(a) Cross-section schematic of the sample and coordinate system.
(b) Image of the device consisting of 8 converging racetracks. 
(c) Hysteresis loop measured by integrating the MOKE contrast over a single racetrack as a function of OOP field $H_\text{z}$ after zero field cooling (blue) and field cooling with $H_\text{FC}\parallel \pm \mathbf{z}$ (green and purple). 
The magnetization vector $\mathbf{m}$ of the Co layer is represented by the double black arrow and the expected alignment of the uncompensated Cr magnetic moments after zero-field cooling by the colored arrows. 
The Cr magnetic moments are aligned opposite to the field cooling direction as they couple antiferromagnetically to the Co magnetization.
(d) Electric wiring and (e) anomalous Hall resistance $R_\text{H}$ as a function of $H_\text{x}$. 
The field $H_\text{x}$ is applied along the x-axis, parallel to the sensing current $j$, and with $\theta=2^{\circ}$ tilt towards the z-axis to promote sharp switching of the magnetization. 
The device shows clear OOP anisotropy with $R_\text{H}$ converging towards 0 at high field as the magnetization is pulled in-plane. 
(f) Hysteresis loops measured by the anomalous Hall effect as a function of $H_\text{x}$ after OOP field cooling ($H_\text{FC}\parallel \pm \mathbf{z}$ with $\theta=90^\circ$, green and purple) and IP field cooling ($H_\text{FC}\parallel \pm \mathbf{x}$ with $\theta=0^\circ$, black and red).
}
\end{figure*}

Magnetic domain wall (DW) motion has been extensively studied in thin film structures to understand magnetization reversal processes \cite{Malozemoff-1979, Thiaville-Book-2006, A512-Metaxas2007, A558-Diaz-PRB2017, A559-Boulle-MatScience2011} and realize memory and logic devices \cite{A560-Marino-IEE1970, A561-Parkin-NatNano2015, A562-Allwood-Sci2005, A421-Luo-Nat2020}. 
Magnetic information can be encoded in DW along continuous strips, the so-called racetrack memories, and an external magnetic field or electric current can precisely displace the DW \cite{A510-Parkin2008, A563-Battarel-IEE-1977, A564-Franken-NatNano2012, A565-Klaui-PRL2005}. 
Out-of-plane (OOP) magnetized ferromagnet/heavy metal (FM/HM) layers are very promising in this respect because their strong perpendicular magnetic anisotropy results in narrow DW with simple Néel or Bloch structure, which can be easily displaced by an OOP external magnetic field \cite{A455-Cayssol2004, A512-Metaxas2007, A446-Je2013, A492-Hrabec2014, A467-Boulle2013-PRL} or current-driven spin-orbit torques \cite{R002-Manchon2019, A424-Thiaville2012, A511-Miron2011, A466-Emori2013, A514-Haazen2013, A469-Ryu2013-NatNano, A426-Martinez2014, A015-Baumgartner2017-NN, A448-Baumgartner2018-APL}. 
The type of DW and their chirality is determined by the Dzyaloshinskii-Moriya interaction (DMI) \cite{A424-Thiaville2012,A566-Heide-PRB2008}, and the response of DW to external stimuli in these systems can be further tuned by interfacial engineering \cite{A567-Torrejon-NatCom2014, A568-Chen-NatCom2013, A478-Lavrjsen2015, A484-Ryu2014-NatCom, A569-Martini-PRA2022}, coupling to additional magnetic layers \cite{A570-Yang-NatNano2015, A461-DeJong2020-PRB}, and electric fields \cite{A571-Chiba-NatCom2012, A572-Schott-JMM2021}.

Typically, reversing the direction of the field or current results in an opposite but symmetric displacement of the DW. Superposing an in-plane (IP) magnetic field breaks this symmetry, which results in different DW velocities depending on whether the DW moves parallel or antiparallel to the IP field direction \cite{A446-Je2013, A492-Hrabec2014, A481-VanatkaJCM-2015}. 
This feature is of particular interest for magnetic logic devices where the IP field can promote or restrict the DW propagation along one IP direction, similar to a magnetic diode. However, variable and selective external fields cannot be easily embedded in miniaturized devices.

Instead of an IP magnetic field, the exchange bias field at the interface between an antiferromagnet (AFM) and a FM \cite{A201-Kiwi2001} can be used to break the symmetry and manipulate the DW dynamics. 
This concept has been successfully used in the context of field-free magnetization switching by spin-orbit torques \cite{A289-Fukami2016, A500-Oh2016, A386-Van2016-NatCom, A573-Krishnaswamy-PRA2020} as well as for field-driven DW motion, for which anti-symmetries in the domain structure and between the ascending and descending branch of the magnetization loop were found in exchange-biased systems \cite{A532-Fitzsimmons2000-PRL, A533-Kirilyuk2002-JAP, A531-McCord2003-JAP, A534-McCord2009-NewJP, A530-Lee2012-JAP, A365-Wu2013, A494-Khan2018, A410-Kuswik2018, A541-Shi2019-JMMM}. 
Moreover, exchange bias can be used to create pinning sites in crossed FM and AFM wires \cite{A405-Albisetti2016-JMMM, A441-Kao2021-AIPadv, A408-Polenciuc2014-APL} and to modify the DW tilt angle \cite{A445-Kim2021-AdvSci}. 
However, a systematic study of how exchange bias affects the DW motion in both field-driven and current-driven experiments is presently lacking.

Here we show that exchange bias in AFM/FM/HM trilayers with perpendicular magnetization can be used to control the direction of motion and velocity of the DW. 
We observe almost unidirectional expansion of domains along the exchange bias field in field-driven DW propagation, and a symmetric (anti-symmetric) modulation of the current-driven DW velocity under positive and negative current when the exchange bias field is parallel (perpendicular) to the current.
Our model system is a Cr$_2$O$_3$/Co/Pt trilayer. 
The Co/Pt subsystem is well-known for its strong perpendicular magnetic anisotropy and efficient field- \cite{A512-Metaxas2007, A446-Je2013, A492-Hrabec2014} and current-driven \cite{A511-Miron2011, A466-Emori2013, A514-Haazen2013, A469-Ryu2013-NatNano, A015-Baumgartner2017-NN} DW dynamics. 
Cr$_2$O$_3$ is an insulating AFM that has been widely used to induce OOP exchange bias in Co/Pt and Co/Pd multilayers \cite{A574-Borisov-PRL2005, A387-He2010-NatMat, A528-Ashida2014-APL, A520-Nozaki2012-IEEE, A582-Shiratsuchi-PRL2012} as well as IP exchange bias in permalloy and CoPt thin films \cite{A465-Dho-PRB2005, A495-Lin-JAP2008, A578-Nozaki-APE2014}. 
Additionally, our study shows that Cr$_2$O$_3$/Co/Pt can be exchange-biased either OOP or IP depending on the field cooling direction. 
Cr$_2$O$_3$ is also of particular interest as the AFM spin order can be efficiently manipulated via the magnetoelectric effect \cite{A574-Borisov-PRL2005, A387-He2010-NatMat, A528-Ashida2014-APL, A583-Binel-EJP2005} and is a prototype material for the realization of magnetoelectric random access memory \cite{A516-Kosub2017-NatCom, A517-Hedrich2021-NatPhy}.

This paper is organized as follows: Section \ref{sec-device-fabriation-setup} describes the sample fabrication and experimental setup. 
Section \ref{sec-IP-and-OOP-exchange-bias} presents the magnetic characterization of Cr$_2$O$_3$/Co/Pt as a function of field cooling direction. Section \ref{sec-field-driven} and \ref{sec-current-driven-EBxyz} report the field- and current-driven DW motion measurements as a function of exchange bias, respectively. 
In Section \ref{sec-current-driven-Hx}, we compare the effect on the DW velocity of exchange bias and an IP external magnetic field, which allows us to estimate the exchange bias field and DMI in our sample. Finally, we summarize our results in Sect. \ref{sec-conclusion}.

\section{\label{sec-device-fabriation-setup} Sample fabrication and experimental setup}

A trilayer of Cr$_2$O$_3$(120~nm)/Co(0.85~nm)/Pt(2~nm) was grown by magnetron sputtering on a Al$_2$O$_3$(0001) substrate. 
The numbers between brackets indicate the thickness of each layer. 
The Ar pressure during the growth was 3 mTorr and the base pressure lower than $5\times 10^{-8}$ mTorr. 
The Cr$_2$O$_3$, Co and Pt layers were all sputtered from nominal composition targets. 
The Cr$_2$O$_3$ layer was grown at 800$^{\circ} \,\text{C}$ and annealed at the same temperature for one hour, then cooled to room temperature for the deposition of the Co and Pt layers. 
The epitaxy of the Cr$_2$O$_3$ was investigated by X-Ray diffraction (XRD) and its thicknesses were measured via X-Ray reflectivity. 
XRD results confirmed the epitaxial growth of Cr$_2$O$_3$ thin film with the (0001) orientation on the Al$_2$O$_3$(0001) substrate.
This corresponds to the typical growth on sapphire (0001) of rhombohedral Cr$_2$O$_3$ with the rhombohedron diagonal pointing out-of-plane (see Appendix~\ref{appendix:epitaxial}). 
No traces of secondary crystal orientations of Cr$_2$O$_3$ could be measured.
In this text we use the conventional hexagonal cell with the 4-axis notation to denote the crystallographic planes.
Additional XRD azimuthal scans performed around the Cr$_2$O$_3$ [0001] direction confirmed the absence of crystal twinning.
Atomic force microscopy analysis showed a smooth and homogenous film surface with root-mean-square roughness smaller than 0.5~nm.
Finally, UV-lithography and reactive ion milling were used to pattern a set of 5~$\mu$m-wide and 50~$\mu$m-long racetracks, as shown in Fig.~\ref{fig:hysteresis-loops}(a,b).

To set the exchange bias direction, the sample was placed on a heating stage and field cooled from $T=320$~K to room temperature in a magnetic field $H_\text{FC} = 1600 \,\text{Oe}$. 
The shift of the magnetic hysteresis loop opposite to the field cooling direction indicates the presence of negative exchange bias, as illustrated in Fig.~\ref{fig:hysteresis-loops}(c)-(f), in agreement with previous reports in similar systems \cite{A582-Shiratsuchi-PRL2012, A584-Lim-JMM2009, A520-Nozaki2012-IEEE}.
The Néel temperature was estimated as the minimum field-cooling temperature required to erase the exchange bias, $T_\text{N}=320$~K (see Appendix~\ref{appendix:coercive-EB-versus-T}). 
The increase of $T_\text{N}$ with respect to the bulk value of 307~K (Ref.~\onlinecite{A465-Dho-PRB2005}) is attributed to the compressive strain of Cr$_2$O$_3$(0001) grown on Al$_2$O$_3$(0001), as discussed in Appendix~\ref{appendix:coercive-EB-versus-T} and Ref.~\onlinecite{A556-Veremchuk2022-NanoSmall}.
Local reorientation of the exchange bias along the racetrack is also possible via current-induced Joule heating. For this purpose, we employed a direct current of $j = 0.5\times 10^{12} \; \text{A/m}^2$ in a field of $H_\text{FC} = 1600 \,\text{Oe}$. 
This technique is interesting for applications where the heat can be generated only locally, which is more energy-efficient than heating the whole sample \cite{A338-Lin2019, A351-Kim2019}. 
Experiments performed for both positive and negative current polarities showed that the the exchange bias is determined by the external magnetic field and that there is no significant effect of the spin-orbit torques, unlike in other AFM systems with higher $T_\text{N}$ (Refs.~\onlinecite{A372-Peng2020, A506-Kang2021, A509-Zhang2021}).

A wide-field magneto-optical Kerr effect microscope (MOKE) in polar configuration was used to image the OOP magnetization component of the Co layer. 
Magnetic contrast was enhanced by taking differential MOKE images, obtained by subtracting from each image a reference image captured in a fully magnetized state. 
Two sets of electromagnets generate the OOP and IP external field. 
Hysteresis loops measured by integrating the MOKE contrast over the racetrack area as a function of the OOP magnetic field $H_\text{z}$ allowed us to evidence the presence of OOP exchange bias after field cooling, as shown in Fig.~\ref{fig:hysteresis-loops}(c). 
For the current-driven DW motion, voltage pulses were injected in a racetrack using a sub-ns pulse generator. 
The impedance matching of the racetrack and pulse generator was achieved by connecting a $50\,\Omega$ resistor in parallel to the device, which reduces back reflection and shortens the pulse rise/fall time. 
We computed the average DW velocity along the racetrack, $v_\text{DW}$, as the total area spanned by the DW displacement divided by the racetrack width and pulse length.

The resistance of the 2~nm-thick Pt layer is expected to be much smaller than that of the 0.85~nm-thick Co layer, hence most of the current flows through the Pt layer. With this assumption, the device resistivity was estimated to be $\rho = 2.15 \times 10^{-7} $~$\Omega$m from a 4-probe measurement of the longitudinal resistance $R = 1075~\Omega$ of a 5-$\mu$m-wide and 50-$\mu$m-long Hall bar device. 
To evidence the presence of IP exchange bias, we measured the anomalous Hall resistance, $R_\text{H} = V_\text{H}/I$ with $V_\text{H}$ the Hall voltage and $I$ the sensing current, of the Co layer as a function of the external field $H_\text{x}$ applied along the x-axis, with a $2^\circ$ tilt towards the z-axis, as shown in Fig.~\ref{fig:hysteresis-loops}(d-f). 
As $R_\text{H}$ is proportional to the OOP component of the magnetization, this type of measurement yields information on the influence of IP exchange bias on the switching of the Co layer.
All the measurements were performed at room temperature.

\section{\label{sec-IP-and-OOP-exchange-bias} Out-of-plane and in-plane exchange bias in \texorpdfstring{C\lowercase{r}$_2$O$_3$/C\lowercase{o}/P\lowercase{t}}{}}

We measured the coercivity, $H_\text{c}$, and exchange bias field, $H_\text{EB}$, for OOP and IP field cooling by recording hysteresis loops as a function of OOP and IP applied fields, respectively. 
For the OOP hysteresis loop, we integrated the MOKE contrast over the racetrack shown in Fig.~\ref{fig:hysteresis-loops}(b) as a function of the OOP field $H_\text{z}$. 
The hysteresis loop of the zero-field cooled device (blue curve) shown in Fig.~\ref{fig:hysteresis-loops}(b) has a coercivity $H_\text{c} = 130\pm3\,\text{Oe}$ and is centered around $H_\text{z}=0\,\text{Oe}$, demonstrating no exchange bias. 
MOKE images show that the reversal occurs by domain nucleation and propagation. 
Upon positive (negative) OOP field cooling, the hysteresis loop (green and purple curves) shifts to negative (positive) field, corresponding to an OOP exchange bias of $H_\text{EB} = -(+) 25 \pm 3\,\text{Oe}$. 
The OOP $H_\text{EB}$ is comparable to previous measurements performed near-room temperature in Cr$_2$O$_3$/Co/Pt \cite{A520-Nozaki2012-IEEE}.
In a simple model, we can represent the AFM spin configuration at the interface pointing opposite to the field cooling direction, as illustrated in Fig.~\ref{fig:hysteresis-loops}(c), which couples antiferromagnetically to the Co spin and shifts the loops \cite{A582-Shiratsuchi-PRL2012}. 
We measured no OOP exchange bias upon IP field cooling.

\begin{figure*}
\centering
\includegraphics[width=0.99\linewidth]{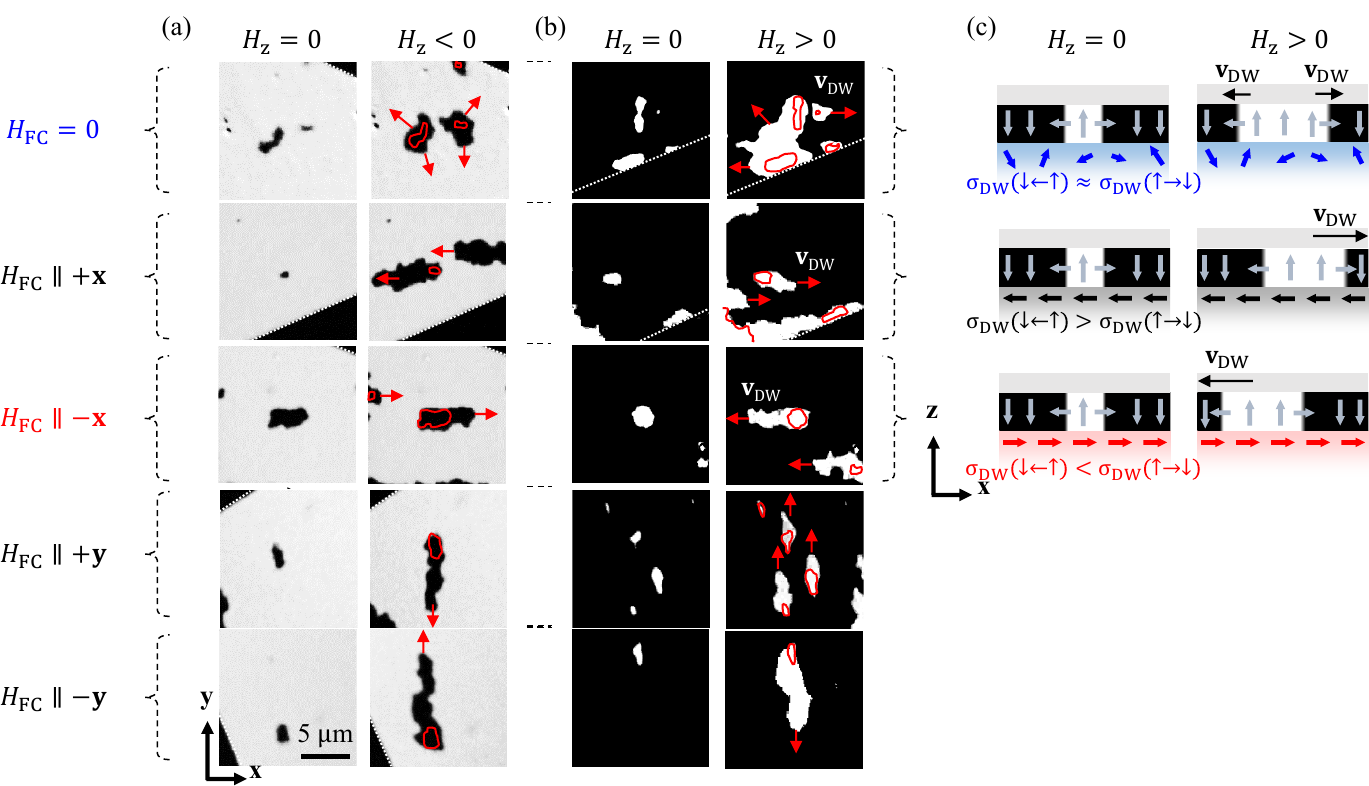}
\caption{\label{fig:field-driven-DW} MOKE images of (a) down domains with $\mathbf{m}\parallel -\mathbf{z}$ (black contrast) and (b) up domains with $\mathbf{z}\parallel+\mathbf{z}$ (white contrast) after nucleation (left column) and expansion (right column) under $H_\text{z} = \mp \mathbf{z}$. 
The field pulse duration is $5~\text{s}$.
The first row shows the zero field-cooled device where the domains expand in all directions, and the next rows show the domains after IP field cooling with $H_\text{FC} = +(-) \mathbf{x} \,\text{and} +(-)\mathbf{y}$. 
The down (up) domains expand preferably antiparallel (parallel) to the direction of $H_\text{FC}$. The red contours indicate the initial shape of the domains before the application of $H_\text{z}$; the red arrows indicate the favored DW motion. 
(c) Schematic of chiral DW and associated domain DW energy, $\sigma_\text{DW}$, for different field cooling directions. The IP exchange bias reduces $\sigma_\text{DW}$ when it is aligned with the internal DW magnetization. 
Upon applying $H_\text{z}$, DW with small energy move faster than DW with large energy (see text for details).}
\end{figure*}

To measure the IP exchange bias, we recorded $R_\text{H}$ as a function of the IP field, $H_\text{x}$, with $2^\circ$ tilting towards the z-axis to allow sharp rotation of the magnetization, as explained above and illustrated in Fig.~\ref{fig:hysteresis-loops}(d).
The hysteresis loop as a function of $H_\text{x}$ of the zero field-cooled device is plotted in Fig.~\ref{fig:hysteresis-loops}(e). 
At $H_\text{x} = 3500\,\text{Oe}$, $R_\text{H}$ changes sign abruptly due to the reversal of the OOP Co magnetization.
This reversal field corresponds to the OOP coercivity as $\cos(2^{\circ})\times 3500 \approx H_\text{c}$.
At fields above $3500\,\text{Oe}$ the Hall resistance decreases and ultimately tends towards zero for $H_\text{x} > 20000\,\text{Oe}$ when the magnetization lies in plane ($m_\text{z} \approx0)$.
Figure~\ref{fig:hysteresis-loops}(f) shows the hysteresis loops as a function of $H_\text{x}$ after positive (negative) OOP field cooling (green and purple curves) and IP field cooling (black and red curves). 
The loops are centered for the OOP field-cooled device, but shifted for the IP field-cooled device by $350 \pm 50 \,\text{Oe}$ opposite to the field cooling direction. 
We cannot directly attribute these shifts to IP exchange bias along the x-axis as its projection along the z-axis is essentially zero, and thus should not contribute to the required OOP switching field. 
However, the shifts demonstrate that the AFM spin configuration at the interface is different upon OOP or IP field cooling. 
Similar shifts were observed when measuring IP hysteresis along the y-axis after field cooling along the y-axis, indicating no IP anisotropy.
Similarly to AFM spin configuration upon OOP field cooling, we suppose that the AFM spins acquire an IP component opposite to the IP field during field cooling, as illustrated in Fig~\ref{fig:hysteresis-loops}(f). 
The ensuing IP exchange bias modifies the energy landscape of the DW and induces anti-symmetric switching behavior as a function of $H_\text{x}$, as discussed in detail in Sect.~\ref{sec-field-driven}. 
We further note that we measured negligible training effects on the exchange bias upon repeated cycling of the applied field (see Appendix~\ref{appendix:coercive-EB-versus-T}).

The shift of the OOP hysteresis loops opposite to the field cooling direction indicates a collinear coupling at the interface between the Cr$_2$O$_3$ and the Co layers, with both AFM and FM spins pointing OOP. 
This behavior is consistent with the epitaxial growth of Cr$_2$O$_3$ with the (0001) orientation that favors the AFM spins alignment perpendicular to the surface \cite{A465-Dho-PRB2005, A582-Shiratsuchi-PRL2012}. 
On the other hand, the shift of the hysteresis loops as a function of $H_\text{x}$ indicates that the AFM spins can also be reoriented IP, while the Co magnetization remains OOP. 
Similar IP canting of the Cr$_2$O$_3$(0001) spins was observed upon coupling to a NiFe layer with IP anisotropy \cite{A465-Dho-PRB2005, A495-Lin-JAP2008}. 

Additionally, the stable IP spin configuration of the Cr$_2$O$_3$ surface coupled to OOP Co spins supports the picture of noncollinear coupling between IP AFM spins and OOP FM spins as suggested in field-free switching of OOP FM layer by spin-orbit torques \cite{A289-Fukami2016, A500-Oh2016, A386-Van2016-NatCom, A573-Krishnaswamy-PRA2020}.

\section{\label{sec-field-driven} Field-driven DW motion}

\begin{figure*}
\centering
\includegraphics[width=0.99\linewidth]{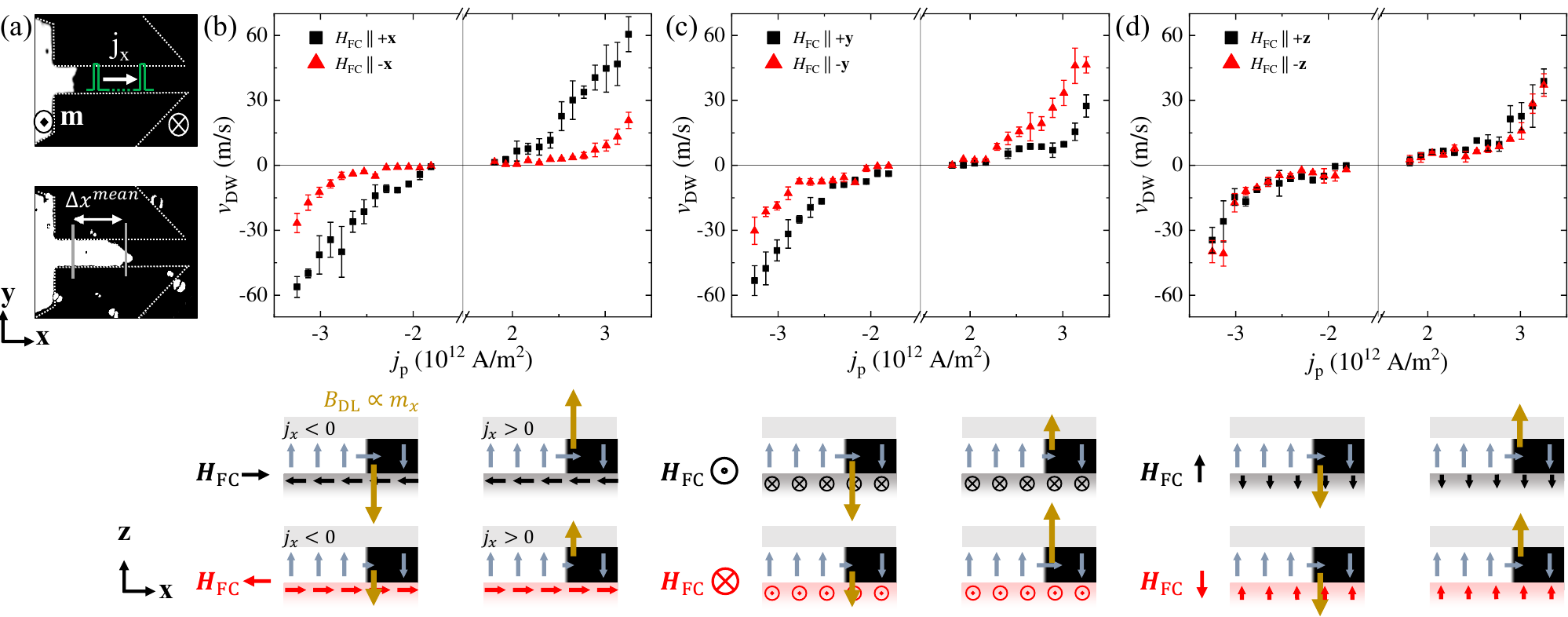}
\caption{\label{fig:current-driven-EBxyz} 
(a) MOKE images of an up-down DW before (top) and after (bottom) applying thirty 4-ns-long pulses of current density $j_\text{p} = 2.5\times 10^{12}\,\text{A/m}^2$ with $10\,\text{Hz}$ repetition rate.
(b-d) $v_\text{DW}$ versus $j_\text{p}$ of up-down DW in the absence of external field after field cooling with $H_\text{FC} \parallel\pm\mathbf{x}$ (b), $H_\text{FC} \parallel\pm\mathbf{y}$ (c), and $H_\text{FC} \parallel\pm\mathbf{z}$ (d). 
For each current density, $v_\text{DW}$ is averaged over four pulse sequences; the error bars represent the standard deviation of each measurement. 
The sketches under each panel exemplify the effect of the exchange bias on $\mathbf{m}_\text{DW}$ (gray arrows) and $B_\text{DL}$ (yellow arrow).}
\end{figure*}

The effect of the exchange bias onto the magnetization is further investigated by inspecting the field-driven DW motion, as reported in Fig.~\ref{fig:field-driven-DW}. 
For isotropic samples, in the absence of an exchange bias, an OOP field acts as a driving force onto the DW magnetization and makes the domains expand with no preferential direction \cite{A455-Cayssol2004, A512-Metaxas2007, A515-Lavrijsen2012, A424-Thiaville2012, A467-Boulle2013-PRL}. 
An in-plane field, $H_\text{IP}$, however, can break this symmetry, because the DW energy, hence the DW velocity $v_\text{DW}$, depend on the relative orientation between the DW magnetization and $H_\text{IP}$, as reported in FM/HM systems \cite{A446-Je2013, A492-Hrabec2014, A481-VanatkaJCM-2015}.

In FM/HM systems, the DMI acts as an effective field $H_\text{DMI}$ on the DW magnetization and stabilizes chiral Néel-type DW \cite{A424-Thiaville2012,A566-Heide-PRB2008}, as illustrated in Fig.~\ref{fig:field-driven-DW}. 
In such a case, for $H_\text{IP}$ applied parallel to $H_\text{DMI}$, the internal DW energy density is given by 
\begin{equation}
\label{eq:DW-energy}
    \sigma_\text{DW} = \sigma_0 + K_\text{D} \, \lambda - \pi \, M_\text{s} \, \lambda \, |H_\text{IP} + H_\text{DMI}|
\end{equation}
with $\sigma_0$ the Bloch-type DW energy density, $K_\text{D}$ the anisotropy energy density of the DW, and $\lambda$ the DW width \cite{A424-Thiaville2012}. 
In the creep regime, when the DW are pinned, a decrease of the DW energy induces an increase of the DW velocity \cite{A455-Cayssol2004, A580-Jeudy-PRB2018, A492-Hrabec2014, A446-Je2013, A513-Shahbazi2019, A594-Diez-APL2020, A593-Gehanne-PRR2020}.
In the flow regime (not reached in our experiment), the DW velocity increase (decrease) is mainly due to the increase (decrease) of the DW width \cite{A581-Jue-PRB2016, A481-VanatkaJCM-2015}.
As a result, applying $H_\text{z}$ alone makes the domains expand in all directions, but adding $H_\text{IP}$ lifts the DW degeneracy and leads to a higher velocity of the DW when $H_\text{IP}$ is parallel to $H_\text{DMI}$.

By analogy with the effect of $H_\text{IP}$ in FM/HM systems, we expect that the IP exchange bias will affect the DW dynamics. 
We examine the expansion of the domains under $H_\text{z}$ by MOKE for different field-cooling directions, as shown in Fig.~\ref{fig:field-driven-DW}(a,b). 
The domains are nucleated using an alternating $H_\text{z}$ from an uniformly magnetized state, and the images are taken before ($H$=0) and after applying $H_\text{z}\lessgtr 0$ for $5\,\text{s}$. 
In the zero-field cooled device the up and down domains tend to expand in all directions, similarly to a non exchange-biased FM/HM system. The rough contour of the domains is a signature of pinning due to the presence of the AFM layer and inhomogeneities in the sample. 
Interestingly, when the sample is IP field cooled with $H_\text{FC} \parallel \pm \mathbf{x}, \pm\mathbf{y}$, the up domain tends to expand along the field-cooling direction, whereas the down domains expand opposite to it. 
A similar result is obtained when $H_\text{IP}$ is applied to a zero-field cooled device. 
This shows that the exchange bias field acts as an effective field on the DW magnetization. Consequently, the internal DW energy density (Eq.~\ref{eq:DW-energy}) can be modified as
\begin{equation}
\label{eq:DW-energy-HEB}
    \sigma_\text{DW} = \sigma_0 + K_\text{D} \, \lambda - \pi \, M_\text{s} \, \lambda \, |H_\text{IP} + H_\text{DMI} + H_\text{EB}|.
\end{equation}

Furthermore, because the DW velocity is higher when the exchange bias is parallel (antiparallel) to the up-down (down-up) DW magnetization and the DW motion is favored parallel to $H_\text{DMI}$ according to Eq.~\ref{eq:DW-energy-HEB}, we deduce that $H_\text{DMI}$ points "to the right" in up-down DW ($\uparrow \rightarrow \downarrow$) and "to the left" in a down-up DW ($\downarrow \leftarrow \uparrow$), giving overall a right-handed chiral DW ($\downarrow \leftarrow \uparrow \rightarrow \downarrow$), as expected for Pt on top of Co (Ref.~\onlinecite{A545-Yang2015-PRL, A449-Jue2016-NatMat}).
This model, in which the IP exchange bias influences $v_\text{DW}$ through the variation of the DW energy density (Eq.~\ref{eq:DW-energy-HEB}), provides a straightforward interpretation of our results. We point out, however, that a quantitative comparison of the DW motion in Cr$_2$O$_3$/Co/Pt trilayers relative to Co/Pt would require including the effects of disorder on the DW velocity \cite{A580-Jeudy-PRB2018}, in particular those due to exchange bias and the Cr$_2$O$_3$ substrate.

\section{\label{sec-current-driven-EBxyz} Current-driven DW motion}

\begin{figure*}
\centering
\includegraphics[width=0.99\linewidth]{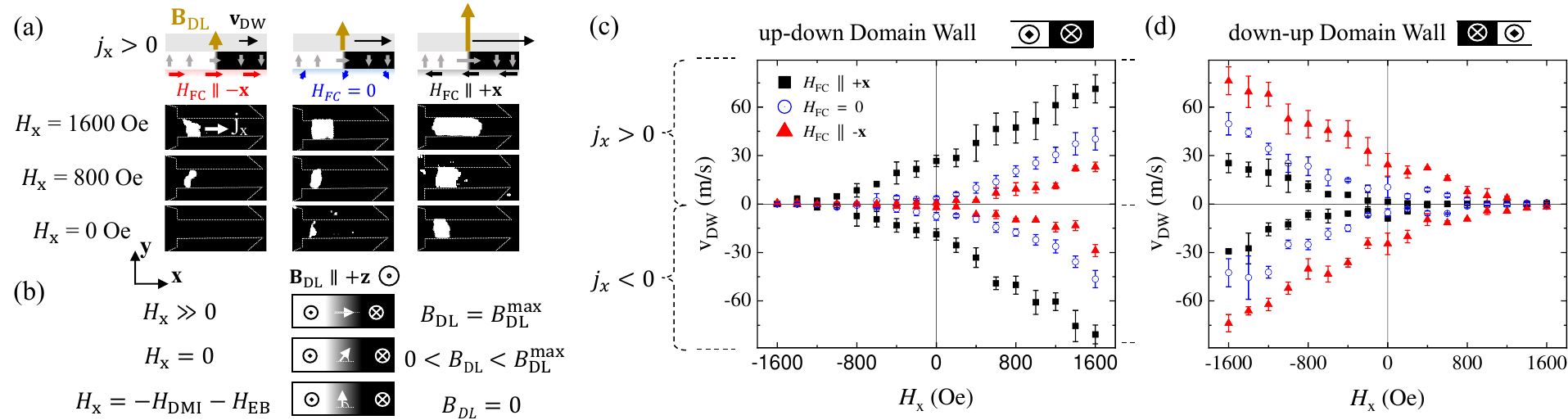}
\caption{\label{fig:current-driven-Hx} 
(a) MOKE images of the displacement of up-down DW after applying thirty 4-ns-long pulses of current density $j_\text{p} = 2.5\times 10^{12}\,\text{A/m}^2$ under $H_\text{x}= 1600, 800, 0 ~\text{Oe}$ (top to bottom rows) for different field-cooling directions.
(b) Schematics illustrating how $\mathbf{m}_\text{DW}$ and $B_\text{DL}$ change depending on $H_\text{x}$, $H_\text{DMI}$ and $H_\text{FC}$.
(c) $v_\text{DW}$ versus $H_\text{x}$ for up-down and (d) down-up DW. 
$H_\text{FC}\parallel + (-) \mathbf{x}$ increases (decreases) the effective IP field $H_\text{x} \pm H_\text{EB}$ acting on the DW, horizontally shifting the curves to larger (smaller) fields with respect to zero field cooling. 
The current density is fixed to $|j_\text{p}|= 2.5\times10^{12} \,\text{A/m}^2$, the pulse length and repetition rate are the same as in Fig.~\ref{fig:current-driven-EBxyz}.}
\end{figure*}

The current-driven DW motion in FM/HM systems is based on the absorption of the spin accumulation at the HM interface, which induces a damping-like spin-orbit-torque (DL-SOT) on the internal DW magnetization, $\mathbf{m}_\text{DW}$, \cite{R002-Manchon2019, A424-Thiaville2012, A466-Emori2013, A514-Haazen2013, A469-Ryu2013-NatNano, A426-Martinez2014, A015-Baumgartner2017-NN, A448-Baumgartner2018-APL}.
For Néel DW with $\mathbf{m}_\text{DW} \parallel \mathbf{j} \parallel \mathbf{x}$, the torque results in an effective easy-axis field, $B_\text{DL}$, which rotates $\mathbf{m}_\text{DW}$ towards $\pm \mathbf{z}$ depending on the relative alignment of $\mathbf{j}$ and $\mathbf{x}$.
This rotation induces the propagation of the DW. The sign of $B_\text{DL}$ changes upon inverting the current direction. 
Additionally, the torque induces a rotation of $\mathbf{m}_\text{DW}$ towards $\mathbf{y}$, which causes tilting of the DW as $H_\text{DMI}$ favors $\mathbf{m}_\text{DW}$ perpendicular to the DW \cite{A467-Boulle2013-PRL, A426-Martinez2014, A448-Baumgartner2018-APL}.

Starting from a DW with $\mathbf{m}_\text{DW} \parallel \mathbf{x}$, as in a racetrack, the application of an IP field $H_\text{x}$ does not exert a torque on the DW magnetization, but rather enhances or opposes the effective field $H_\text{DMI}$ that stabilizes the Néel DW configuration along $\mathbf{x}$. 
Because $B_\text{DL}$ is maximum when $\mathbf{m}_\text{DW} \parallel \pm \mathbf{x}$, $H_\text{x}$ increases or decreases the current-driven $v_\text{DW}$. 
The change of $v_\text{DW}$ for a fixed $H_\text{x}$ is anti-symmetric with respect to current inversion.
A field $H_\text{y}$, on the other hand, either supports or opposes the rotation of $\mathbf{m}_\text{DW}$ towards $\mathbf{y}$. As a consequence, $H_\text{y}$ results in an anti-symmetric variation of $v_\text{DW}$ depending on the current direction \cite{A469-Ryu2013-NatNano,A426-Martinez2014}.

Based on the results of Sect.~\ref{sec-field-driven}, we expect that the IP exchange bias should produce similar effects on $v_\text{DW}$ as those described above for the IP fields $H_\text{x,y}$. 
We thus investigate the current-driven DW motion in Cr$_2$O$_3$/Co/Pt racetracks for different directions of the IP exchange bias. We emphasize here the importance of minimizing Joule heating in our samples because of the relatively low Néel temperature of Cr$_2$O$_3$. 
We achieve this by utilizing short current pulses of limited amplitude. We verified that the exchange bias vanishes when applying pulses longer than $10\,\text{ns}$ with current density $j_\text{p} > 2 \times 10^{12} \,\text{A/m}^2$ and cannot be retrieved without performing another field cooling (see in Fig.~\ref{fig:appendix:pulse-length}).

Images of an up-down DW before and after applying a series of current pulses in the absence of an external field are shown in Fig.~\ref{fig:current-driven-EBxyz}(a). 
The DW was initially positioned in the racetrack using a combination of $H_\text{x}$ and $H_\text{z}$ external fields.
The DW tends to be pinned at defects and at the edges of the racetrack, which deforms the DW boundary. 
Figure~\ref{fig:current-driven-EBxyz}(b-d) shows $v_\text{DW}$ as function of $j_\text{p}$, for $H_\text{FC} \parallel\pm\mathbf{x}$, $\pm\mathbf{y}$, and $\pm\mathbf{z}$ with no applied external field. 
All the curves are characterized by a finite critical current for DW motion, a gradual increase of $v_\textrm{DW}$ corresponding to the creep regime, and a curvature change representing the depinning threshold followed by a linear region in which $v_\textrm{DW}$ increases proportionally to $j_\text{p}$, as expected in the flow regime. These curves are typical of spin-orbit torque-driven Néel DW motion \cite{A511-Miron2011, A466-Emori2013, A469-Ryu2013-NatNano}.

However, we find substantial differences in the curves as a function of the field cooling direction. $H_\text{FC} \parallel + (-) \mathbf{x}$ decreases (increases) the depinning threshold along both the positive and negative current direction [Fig.~\ref{fig:current-driven-EBxyz}(b)]. 
For up-down DW, $H_\text{FC} \parallel + (-) \mathbf{y}$ increases (decreases) the depinning threshold for positive current, and decreases (increases) it for negative current [Fig.~\ref{fig:current-driven-EBxyz}(c)].
The opposite effects are observed for down-up DW.
Hence, the effects of the IP field cooling on $v_\text{DW}$ have the same symmetries as the effects of $H_\text{x}$ and $H_\text{y}$ in FM/HM system discussed above. The DW velocity versus current characteristics with $H_\text{FC} \parallel \mathbf{y}$ is that of a DW diode, a useful component of magnetic DW logic circuits \cite{A421-Luo-Nat2020, A589-LuoZhao-PRA2021}. Using the exchange bias, this functionality is obtained without a specific design of the racetrack \cite{A587-Allwood-APL2004, A588-Bryan-APL2007} or applying external magnetic field.
On the other hand, $H_\text{FC}\parallel\pm\mathbf{z}$ does not affect $v_\text{DW}$ within the accuracy of our measurements [Fig.~\ref{fig:current-driven-EBxyz}(d)].

Overall, we find that the exchange bias changes the low current regime by shifting the depinning threshold, whereas the DW mobility in the high current regime appear to be less affected. The exchange bias field significantly increases or decreases $v_\text{DW}$ in a symmetric or anti-symmetric way depending on the field-cooling direction, which is particularly interesting for controlling the DW motion in the absence of external fields.

\section{\label{sec-current-driven-Hx} Estimate of the IP exchange bias field by current-driven DW motion}

The magnitude of the effective fields $H_\text{EB}$ and $H_\text{DMI}$ acting on $\mathbf{m}_\text{DW}$ can be estimated by measuring $v_\text{DW}$ versus $H_\text{x}$ and finding the field at which $v_\text{DW} = 0$. The reasoning here is similar to that applied to FM/HM systems in the absence of exchange bias \cite{A469-Ryu2013-NatNano, A425-Emori2014-PRB}. In these systems, when $H_\text{x} + H_\text{DMI}= 0$ the DW changes from Néel to Bloch-type, as the latter is the favored DW configuration for a thin film with OOP magnetization in the absence of DMI. This in turn stops the DW motion because $B_\text{DL}=0$ when $\mathbf{m}_\text{DW} \parallel \pm \mathbf{y}$ \cite{A424-Thiaville2012}. In our exchange-biased Co layer the same occurs when $H_\text{x} + H_\text{EB} + H_\text{DMI}= 0$, as illustrated in Fig.~\ref{fig:current-driven-Hx}(a,b).

Figure~\ref{fig:current-driven-Hx}(c,d) shows $v_\text{DW}$ versus $H_\text{x}$ measured for up-down DW (c), and down-up DW (d), for positive and negative current (upper and lower part of the panels) at fixed current density.
We observe that $v_\text{DW}$ of the up-down (down-up) DW increases for positive (negative) $H_\text{x}$ and both DW move along the current direction at $H_\text{x}=0$.
This behavior is characteristic of right-handed chiral Néel DW \cite{A466-Emori2013, A469-Ryu2013-NatNano, A425-Emori2014-PRB}, which confirms the positive (negative) sign of $H_\text{DMI}$ along $\mathbf{x}$ for up-down (down-up) DW as determined in Sect.~\ref{sec-field-driven}.
$H_\text{FC} \parallel \pm \mathbf{x}$ shifts the curves to lower (higher) $H_\text{x}$ with respect to zero field cooling, and the shift is symmetric for positive and negative current (upper and lower part of the graphs respectively).
These shifts correspond to an effective IP field acting on the DW and can be attributed to the IP exchange bias, $H_\text{EB} \approx 800\,\text{Oe}$.
At zero external field, the average $v_\text{DW}$ increases by a factor ten, from $|2.4| ~\text{m/s}$ to $|23.6|~\text{m/s}$, when $H_\text{EB}$ is along or opposite to $H_\text{DMI}$.
By considering the field interval over which $v_\text{DW} =0$ in Fig.~\ref{fig:current-driven-Hx}(c,d) for the DW in which the DMI and exchange bias field oppose each other, we estimate $H_\text{DMI}\approx 1700$~Oe, which is consistent with reported values in Co/Pt systems \cite{A492-Hrabec2014, A446-Je2013, A488-Franken2015-SciRep}.

We note that $H_\text{EB}$ is larger than the shift of the hysteresis loops obtained by the anomalous Hall measurements of the OOP magnetization as a function of a tilted IP field, which amounts to $350 \pm 50 \,\text{Oe}$ [see Fig.~\ref{fig:hysteresis-loops}(f)].
This is not surprising because the shifted loops in Fig.~\ref{fig:hysteresis-loops}(f) reflect the influence of the IP exchange bias on DW nucleation, whereas the measurements in Fig.~\ref{fig:current-driven-Hx} reflect the influence of exchange bias on $\mathbf{m}_\text{DW}$ and DW motion. 
More surprising is the fact that the IP exchange bias is about one order of magnitude larger than the OOP exchange bias estimated by the shift of the hysteresis loops as a function of $H_\text{z}$. 
This is unexpected given the OOP anisotropy of the Co layer and of bulk Cr$_2$O$_3$(0001). 
We propose two different interpretations of this result. One possibility is that the model used to estimate the IP exchange bias cannot capture the full complexity of the system because it assumes a variation of $v_\text{DW}$ solely based on the variation of the DW energy density, as exemplified by Eq.~\ref{eq:DW-energy-HEB}. 
Another possibility is that the in-plane compressive strain of Cr$_2$O$_3$(0001) grown on Al$_2$O$_3$(0001) favors the transition from easy-axis OOP anisotropy of bulk unstrained Cr$_2$O$_3$ to easy-plane IP anisotropy, as theoretically predicted \cite{A626-Mu-PRM2019}. The latter effect is supported by the $0.6\%$ elongation of the Cr$_2$O$_3$ crystal structure along the [0001] direction measured by XRD (see Appendix~\ref{appendix:epitaxial}).

We observe that $v_\text{DW}$ has a nonlinear dependence on $H_\text{x}$ close to the field at which $H_\text{x} + H_\text{EB} + H_\text{DMI}= 0$, unlike the linear dependence that is usually reported or assumed for FM/HM systems \cite{A469-Ryu2013-NatNano, A484-Ryu2014-NatCom, A426-Martinez2014, A485-Torrejon-MagnNano2020}.
We attribute this behavior to the gradual change of the DW from Néel to Bloch-type [Fig.~\ref{fig:current-driven-Hx}(b)]. The change starts when $H_\text{x} + H_\text{EB} + H_\text{DMI}| \leq \frac{2}{\pi}\,H_\text{K}$ \cite{A425-Emori2014-PRB}, where $H_\text{K} = \frac{\ln(2)\,t\,\mu_0\,M_\text{s}}{\pi \Delta}$ is the shape anisotropy field that favors Bloch DW, $t$ is the FM thickness, $M_\text{s}$ the saturation magnetization, and $\Delta$ the DW width.
Taking $t = 0.85\,\text{nm}$, $M_\text{s} = 7.2 \times 10^5\,\text{A/m}$ from SQUID measurements, the perpendicular anisotropy field $\mu_0\,H_\text{K}^\perp = 1.5 \,\text{T}$ from Fig.~\ref{fig:hysteresis-loops}(e), the effective anisotropy energy $K_\text{eff}= \frac{\mu_0\,H_\text{K}^\perp\,M_\text{s}}{2}= 5.4\times 10^5 \,\text{J/m}^3$, and the exchange stiffness $A = 15 \,\text{pJ/m}$ (Ref.~\onlinecite{A549-Girt2011-JAP}), we estimate $\Delta =\sqrt{A/K_\text{eff}} = 5.3 \,\text{nm}$. 
Combining these values we find an estimated DW shape anisotropy $H_\text{K} \approx 350 \,\text{Oe}$ which is in agreement with the range of fields over which $v_\text{DW}$ changes nonlinearly starting from zero [Fig.~\ref{fig:current-driven-Hx}(c,d)].


\section{\label{sec-conclusion}Conclusions}

In summary, we studied the effect of exchange bias on the field- and current-driven DW motion of an AFM/FM/HM trilayer with perpendicular magnetic anisotropy and Néel DW stabilized by the DMI.
We found that the exchange bias field in Cr$_2$O$_3$(0001)/Co/Pt can be set either OOP or IP depending on the field cooling direction, while the Co magnetization remains OOP.
The possibility to induce IP exchange bias in a system with perpendicular magnetic anisotropy allows for tailoring the DW velocity and introduce directional asymmetry in the DW dynamics.
Upon applying an OOP magnetic field, we find that an IP exchange bias induces almost unidirectional expansion of the DW with internal magnetization parallel to the exchange bias field.
Upon applying a current, an IP exchange bias significantly offsets the depinning threshold of the DW, leading to a manifold increase (decrease) of the DW velocity when the exchange bias field is set along the current direction parallel (antiparallel) to the DW magnetization. 
If the exchange bias field is set perpendicular to the current direction, the DW velocity increases (decreases) when the bias field opposes (favors) the tilt of the DW magnetization away from the current direction.
To a first approximation, our results show that the IP exchange bias field adds to the effective DMI field and external IP field to determine the DW motion driven by an OOP field or spin-orbit torques.
Exchange bias can thus be used to replace an external field to set a preferential direction of field-driven and current-driven DW motion in perpendicular AFM/FM/HM systems.
By exploiting local current-induced heating, the exchange bias field can in principle be set independently on different racetracks, which is of interest to offset or harmonize the DW motion in magnetic memory and logic devices.

\section*{Acknowledgements}
We are grateful to G. Krishnaswamy and A. Hrabec for insightful discussions. 
This work was funded by the Swiss National Science Foundation (Grant No. PZ00P2-179944 and 200020-200465). P.N. acknowledges support from the ETH Zurich Postdoctoral Fellowship Program 19-2 FEL-61.
S.V. acknowledges support by the Spanish Ministry of Science and Innovation (Grant No. PID2021-122980OA-C53) and by the Comunidad de Madrid through the  Atraccion de Talento program (Grant No. 2020-T1/IND-20041).

\section*{Data availability}
The data that support the findings of this study have been deposited in the Research Collection database of the ETH Zurich and are available from https://doi.org/10.3929/ethz-b-000545102.

\appendix

\section{\label{appendix:epitaxial} Epitaxial growth of \texorpdfstring{C\lowercase{r}$_2$O$_3$}{} on  \texorpdfstring{A\lowercase{l}$_2$O$_3$}{}(0001)}

\begin{figure}
    \centering
    \includegraphics[width=0.99\linewidth]{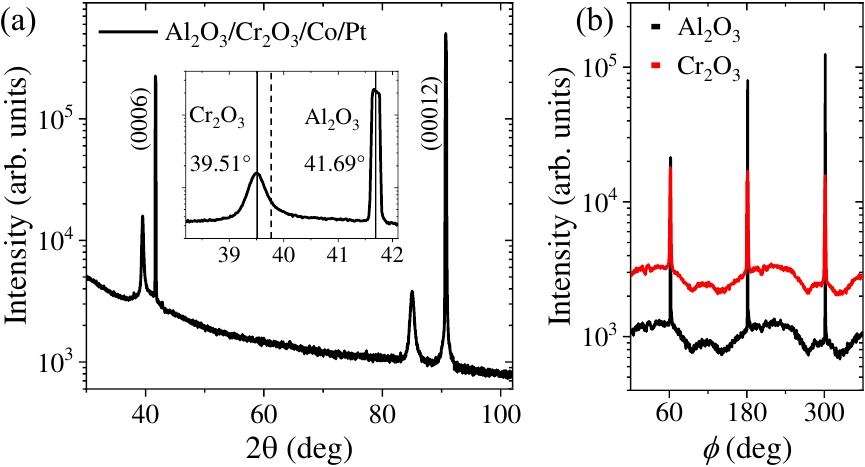}
    \caption{(a) XRD $2\theta$-scan of Cr$_2$O$_3$/Co/Pt deposited on $\alpha$-Al$_2$O$_3$(0001). The inset shows the enlarged scan around the peak corresponding to the (0006) plane of Cr$_2$O$_3$. The dashed line indicates the expected peak position of the (0006) plane of unstrained single-crystalline Cr$_2$O$_3$ (Ref.~\onlinecite{A556-Veremchuk2022-NanoSmall}). (b) Azimuthal XRD scan around the $[0001]$ direction of Cr$_2$O$_3$ (aligned to the $[0001]$ direction of Al$_2$O$_3$) showing the reflexes of the $(10\bar{1}4)$ planes for both Al$_2$O$_3$ (red) and Cr$_2$O$_3$ (black).}
    \label{fig:appendix:2theta-and-phi-scan}
\end{figure}

We characterized the crystal structure of Cr$_2$O$_3$ by XRD investigating a trilayer of Cr$_2$O$_3$(120 nm)/Co(1.1 nm)/Pt(2 nm) deposited on Al$_2$O$_3$(0001) in the same conditions as the sample used to examine DW motion. 
Figure~\ref{fig:appendix:2theta-and-phi-scan}(a) shows the XRD $2\theta$-scan.
The peaks at $41.69^\circ$ and $90.73^\circ$ correspond to the (0006) and (00012) planes of Al$_2$O$_3$, respectively \cite{A630-Yang-APL2011}. 
The peaks at $39.51^\circ$ and $85.04^\circ$ correspond to the (0006) and (00012) planes of Cr$_2$O$_3$ respectively (Ref.~\onlinecite{A629-McMurdie1986}). 
The absence of major additional peaks confirms the epitaxy of Cr$_2$O$_3$ film  growing exclusively with the (0001) orientation. 
Further, the peak of the (0006) plane of Cr$_2$O$_3$ at $39.51^\circ$ is shifted with respect to the peak position of unstrained single-crystalline Cr$_2$O$_3$ at $39.77^\circ$ (Ref.~\onlinecite{A556-Veremchuk2022-NanoSmall}, dashed line in the inset).
This shift indicates that the Cr$_2$O$_3$ unit cell is elongated along the out-of-plane direction. 
The out-of-plane lattice parameter $c_\textrm{exp}$ of the film is measured using Bragg’s law of diffraction $n\lambda = 2d\sin(\theta)$, where $n$ is the diffraction order, $\lambda=1.541$~\AA~the wavelength of the Cu $K_{\alpha 1}$ radiation and $\theta$ the Bragg angle. 
We estimate $c_\textrm{exp}=13.671~\textrm{\AA}$ of the deposited Cr$_2$O$_3$ film (from the (0006) plane at $2\theta = 39.51^\circ$) which differs from $c_\textrm{0}=13.593$~\AA\ of unstrained single-crystal Cr$_2$O$_3$ (Ref.~\onlinecite{A556-Veremchuk2022-NanoSmall}). 
The unit cell is then elongated by $(c_\textrm{exp}-c_\textrm{0})/c_\textrm{0}=0.6\%$ with respect to unstrained crystal.

We attribute this deformation to the in-plane lattice mismatch between the Cr$_2$O$_3$ epitaxial film and the Al$_2$O$_3$ substrate since the latter has an in-plane lattice parameter $4\%$ smaller than the former.
The Cr$_2$O$_3$ unit cell then exhibits an in-plane compressive strain, and consequently, also an out-of-plane tensile strain \cite{A556-Veremchuk2022-NanoSmall}.  
As discussed in detail in Appendix~\ref{appendix:coercive-EB-versus-T}, this lattice deformation is expected to increase the N\'eel temperature $T_\text{N}$ with respected to unstrained single-crystalline Cr$_2$O$_3$.

\textit{Absence of twinning in Cr$_2$O$_3$.}
Figure~\ref{fig:appendix:2theta-and-phi-scan}(b) shows the azimuthal XRD scan around the $[0001]$ direction of Cr$_2$O$_3$ (aligned to the $[0001]$ direction of Al$_2$O$_3$) showing the reflexes of the $(10\bar{1}4)$ planes for both Al$_2$O$_3$ (red) and Cr$_2$O$_3$ (black).
The patterns of Al$_2$O$_3$ and Cr$_2$O$_3$ have the same three-fold in-plane symmetry, confirming the growth of twin-free Cr$_2$O$_3$ epitaxial films \cite{A631-Vu-ScienceReport2020}. 
Furthermore, the alignment of the patterns shows that the in-plane orientation of the rhombohedral Cr$_2$O$_3$ lattice is aligned with the Al$_2$O$_3$ lattice \cite{A630-Yang-APL2011}. 

\section{\label{appendix:coercive-EB-versus-T} Exchange bias vs temperature and estimate of $T_\text{N}$}

\begin{figure}[b]
    \centering
    \includegraphics[width=0.99\linewidth]{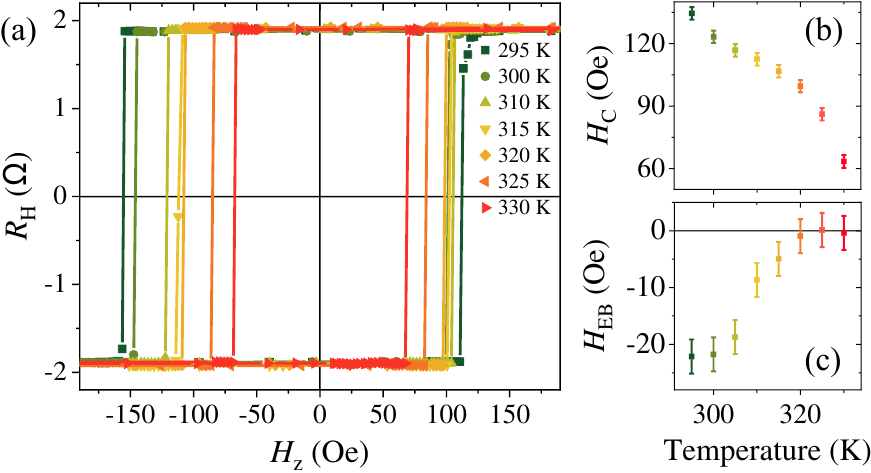}
    \caption{(a) Hysteresis loops at different temperatures. The exchange bias is set once before starting the full set of hysteresis loops by field cooling the sample from 320 to 295~K in an out-of-plane field $H_\textrm{FC}=1600$~Oe. (b) Coercive field and (c) exchange bias field estimated from the loops shown in (a).}
    \label{fig:appendix:heating}
\end{figure}

To estimate $T_\text{N}$ we measured the anomalous resistance $R_\textrm{H}$ as a function of the out-of-plane magnetic field $H_\textrm{Z}$ after setting the exchange bias by field cooling the sample from 320 to 295~K in an out-of-plane field $H_\textrm{FC}=1600$~Oe and recording hysteresis loops at different temperatures up to 330~K. The device is a single Hall cross of width 5~$\mu$m. The hysteresis loop, the coercive field $H_\textrm{c}$ and the exchange bias field $H_\textrm{EB}$ as a function of temperature are presented in Fig.~\ref{fig:appendix:heating}(a-c). The coercive field and the exchange bias decrease gradually with increasing temperature. The coercive field is $H_\textrm{C}=133\pm 5$~Oe at $T=295$~K and decreases to $H_\textrm{C}=67\pm 5$~Oe at $T=320$~K. The exchange bias is $H_\textrm{EB}= 22\pm5$~Oe at $T=295$~K and vanishes at $T\geq320$~K, indicating that $T_\textrm{N}\approx 320$~K. 

\begin{figure}[th]
    \centering
    \includegraphics[width=0.99\linewidth]{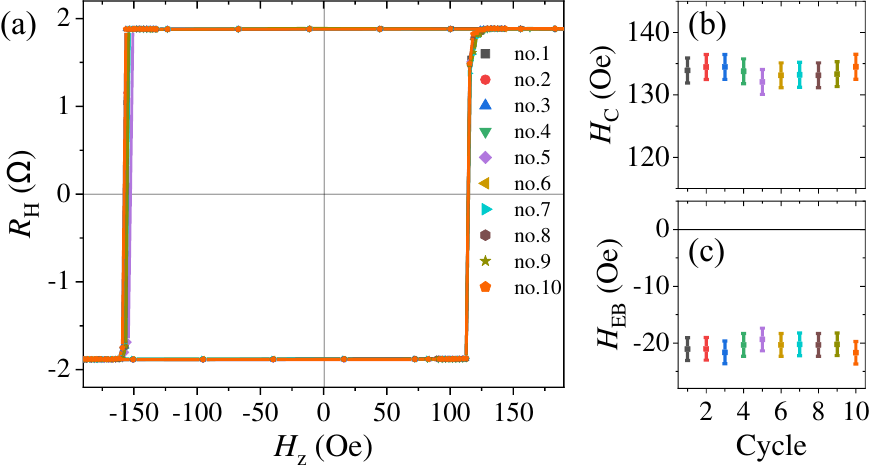}
    \caption{(a) Ten consecutive magnetization loops recorded by measuring the anomalous Hall resistance after field cooling the sample from 320 to 295~K in an out-of-plane field $H_\textrm{FC}=1600$~Oe. (b) Coercive field and (c) exchange bias field estimated from the loops shown in (a).}
    \label{fig:appendix:training}
\end{figure}

\textit{Absence of training effects.}
Figure~\ref{fig:appendix:training}(a) shows ten consecutive magnetization cycles recorded at room temperature after field cooling the sample from $320$~K in an out-of-plane field $H_\textrm{FC}=1600$~Oe. The coercive field and exchange bias are reported in Fig.~\ref{fig:appendix:training}(b-c). These values do not vary significantly from cycle to cycle, indicating the absence of a training effect. 

\begin{figure}[t]
    \centering
    \includegraphics[width=0.99\linewidth]{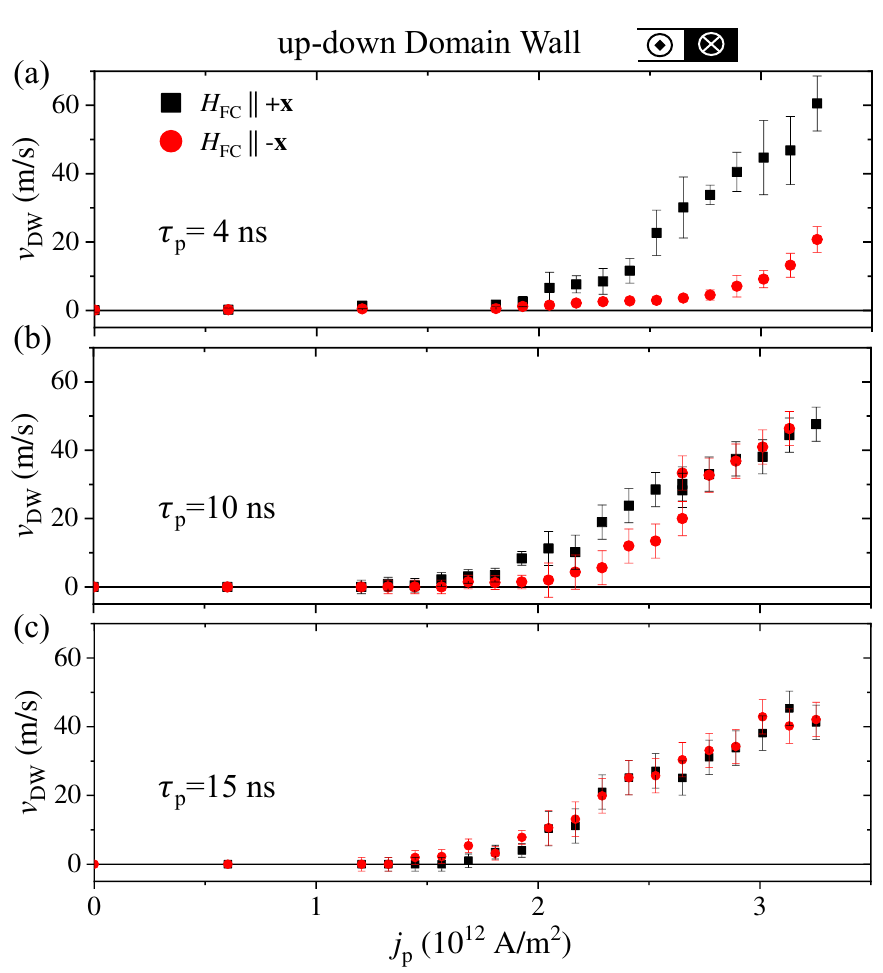}
    \caption{$v_\textrm{DW}$ versus $j_\textrm{p}$ for up-down DW after applying thirty pulses of length (a) $\tau_\textrm{p} = 4$~ns (b) 10~ns and (c) 15~ns with 10~Hz repetition rate in the absence of external field after field cooling with $H_\textrm{FC}\parallel\pm \mathbf{x}$. For each current density, $v_\textrm{DW}$ is averaged over four pulse sequences; the error bars represent the standard deviation of each measurement.}
    \label{fig:appendix:pulse-length}
\end{figure}

\textit{Increase of the Néel temperature due to strain.}
The value of $T_\text{N}$ deduced from the vanishing of exchange bias is higher than the bulk N\'eel temperature of Cr$_2$O$_3$. In line with theoretical calculations, we attribute the increase of $T_\text{N}$ to the residual strain of Cr$_2$O$_3$ thin films grown on Al$_2$O$_3$(0001). Kota et al. \cite{A625-Kota-APE2013} calculated that $T_\textrm{N}$ increases by $20\%$ for a $5\%$ increase of the ratio $c_\textrm{exp}/a_\textrm{exp}$ relative to the unstrained ratio $c_\textrm{0}/a_\textrm{0}$, where $a$ is the in-plane lattice parameter and $c$ the out-of-plane lattice parameter of Cr$_2$O$_3$. 

To estimate the experimental ratio, we use the out-of-plane lattice parameter $c_\textrm{exp}=13.671~\textrm{\AA}$ calculated from the XRD $2\theta$-scan presented in Fig.~\ref{fig:appendix:2theta-and-phi-scan}(a) and assume the in-plane lattice parameter $a_\textrm{exp}=a_\textrm{0}=4.959$~\AA . The assumption of using the unstrained value of $a$ is partially motivated by observations made in Ref.~\onlinecite{A556-Veremchuk2022-NanoSmall}, where TEM measurement performed on a 250-nm-thick Cr$_2$O$_3$ grown on $\alpha$-Al$_2$O$_3$ showed in-plane relaxation of the Cr$_2$O$_3$ lattice. Because of the reduced thickness of our films compared to Ref.~\onlinecite{A556-Veremchuk2022-NanoSmall}, however, we expect that $a_\textrm{exp} \lesssim a_\textrm{0}$ due to the compressive strain imposed by the Al$_2$O$_3$ substrate.

We then find $c_\textrm{exp}/a_\textrm{0}=2.76$, which is about $0.6\%$ larger than the corresponding ratio in the unstrained crystal, $c_\textrm{0}/a_\textrm{0}=2.74$. According to the linear relation between strain and change of $T_\textrm{N}$ from Ref.~\onlinecite{A625-Kota-APE2013}, the out-of-plane tensile strain corresponds to a $2.3\%$ increase of $T_\textrm{N}$. Considering the bulk N\'eel temperature of 307~K (Ref.~\onlinecite{A582-Shiratsuchi-PRL2012}), this gives an estimated N\'eel temperature of 314~K. Because $a_\textrm{exp}$ is likely smaller than $a_\textrm{0}$ in our films, this estimate provides a lower limit for the expected increase of $T_\textrm{N}$ due to strain, which is in good agreement with $T_\textrm{N}\approx 320$~K obtained from the measurements reported in Fig.~\ref{fig:appendix:heating}(c).

\section{\label{appendix:pulse-length} Current-induced domain wall velocity for different pulse lengths}

Figure~\ref{fig:appendix:pulse-length} shows the DW velocity $v_\textrm{DW}$ versus $j_\textrm{p}$ for different pulse lengths (a) $\tau_\textrm{p}=4$~ns, (b) $\tau_\textrm{p}=10$~ns and (c) $\tau_\textrm{p}=15$~ns. All the curves are characterized by a finite critical current for DW motion, a gradual increase of $v_\textrm{DW}$ in the creep regime and a depinning threshold preceding the flow regime in which $v_\textrm{DW}$ increases linearly with $j_\text{p}$, as discussed in Sect.~\ref{sec-current-driven-EBxyz}. 

We observe that the curves for $H_\text{FC} \parallel \pm \mathbf{x}$ measured using the shorter pulses with $\tau_\textrm{p} = 4$~ns are well separated due to the exchange bias favoring DW propagation along the field cooling direction (see Sect.~\ref{sec-current-driven-EBxyz}). For $\tau_\textrm{p} = 10$~ns, the curves superimpose at $j_\textrm{p}>2.5\cdot 10^{12}$~A/m$^2$ and for $\tau_\textrm{p} = 15$~ns the curves overlap in the entire range of current density. This behavior indicates that the device temperature remains below $T_\text{N}$ at intermediate pulse lengths and current density, but exceeds $T_\textrm{N}$ for longer pulse length and higher currents due to Joule heating.

Additionally, we observe that for $H_\textrm{FC}\parallel+\mathbf{x}$ the depinning threshold, estimated as the current density above the creep regime where the curvature changes \cite{A595-Pardo-PRB2019}, decreases upon increasing $\tau_p$, from $j_\textrm{d} (\tau_\textrm{p}=4~\textrm{ns})\approx2.2\cdot 10^{12}$~A/m$^2$ to $j_\textrm{d} (\tau_\textrm{p}=10~\textrm{ns})\approx 2\cdot 10^{12}$~A/m$^2$ and $j_\textrm{d} (\tau_\textrm{p}=15~\textrm{ns})\approx 1.8\cdot 10^{12}$~A/m$^2$. This decrease is expected based on the higher temperature reached by the sample for longer pulses, which favors the thermally-activated depinning of DW.

\bibliography{references}

\end{document}